\documentclass[manuscript]{aastex}
\usepackage{amsmath}
\usepackage{natbib}
\shorttitle{K-shell photoionization of Ni ions}              
\shortauthors{Witthoeft et~al.}

\begin{document}

\title{K-shell photoionization of Nickel ions using $R$-matrix}
\author{M.~C. Witthoeft\altaffilmark{1,2}, M.~A. Bautista\altaffilmark{3}, J. Garc\'ia\altaffilmark{3,2}, T.~R. Kallman\altaffilmark{2}}
\author{C. Mendoza\altaffilmark{4}, P. Palmeri\altaffilmark{5}, and P. Quinet\altaffilmark{5,6}}

\altaffiltext{1}{Department of Astronomy, University of Maryland, College Park, MD, USA}
\altaffiltext{2}{NASA/Goddard Space Flight Center, Code 662, Greenbelt, MD, USA}
\altaffiltext{3}{Department of Physics, Western Michigan University, Kalamazoo, MI, USA}
\altaffiltext{4}{Centro de F\'{\i}sica, Instituto Venezolano de Investigaciones Cient\'{\i}ficas (IVIC), Caracas, Venezuela}
\altaffiltext{5}{Astrophysique et Spectroscopie, Universit\'{e} de Mons - UMONS, B-7000 Mons, Belgium}
\altaffiltext{6}{INPAS, Universit\'e de Li\`ege, B-4000 Li\`ege, Belgium}


\begin{abstract} 
We present $R$-matrix calculations of photoabsorption and photoionization cross
sections across the K edge of the Li-like to Ca-like ions stages of Ni.
Level-resolved, Breit-Pauli calculations were performed for the Li-like to 
Na-like stages. 
Term-resolved calculations, which include the mass-velocity and Darwin 
relativistic corrections, were performed for the Mg-like to Ca-like ion stages.
This data set is extended up to Fe-like Ni using the distorted wave 
approximation as implemented by {\sc autostructure}.
The $R$-matrix calculations include the effects of radiative and Auger dampings
by means of an optical potential.
The damping processes affect the absorption resonances converging to the 
K thresholds causing them to display symmetric profiles of constant width that
smear the otherwise sharp edge at the K-shell photoionization threshold. 
These data are important for the modeling of features found in photoionized
plasmas.
\end{abstract}

\keywords{atomic data --
atomic processes -- line formation -- X-rays: spectroscopy}

\normalsize   


\section{Introduction}

The K lines and edges of iron and nickel are important diagnostics in X-ray 
astronomy.  
They are  in a relatively unconfused part of the X-ray spectrum 
($\sim 6$--10 keV), and they are formed over a wide range of physical 
conditions.  This can range 
from cold (by X-ray standards) neutral gas at temperatures $\leq 10^4$ K, to 
highly ionized gas near $10^8$ K.  
Their energy and shape give information about the ionization and kinematics 
in the gas.
The strengths of emission features, such as K fluorescence lines, constrain the
amount of emitting gas and the elemental abundances, while the strengths of 
absorption edges constrain the column densities along the line of sight to the
background X-ray source.
Iron K lines and edges have been studied by many existing and past X-ray
astronomy instruments.  
They have provided insights into diverse topics such as elemental abundances in
clusters of galaxies \citep{serletal77}, the spin of black holes 
\citep{brenetal11,mccletal11}, and the geometry of stellar flares 
\citep{testetal08}.
Nickel features have not yet been used to the same extent, owing to the fact 
that they are weaker by a factor $\sim$10 from the typical Ni/Fe abundance 
ratio, and to the fact that many detectors have collecting areas which 
decrease rapidly above 7 keV.
Nonetheless, Nickel has been detected in systems ranging from black hole X-ray 
binaries \citep{milletal06,kalletal09}, clusters of galaxies \citep{tamuetal09},
the center of our galaxy \citep{koya11}, and supernova remnants \citep{tama10}.
Further impetus for calculations of the spectra of nickel comes from the 
{\em Astro-H} satellite, which will result in the most sensitive observations 
to date of objects at energies containing K features of nickel.  This is 
because of both the relatively large sensitivity at the energy of the nickel K 
lines and edges (7.2 -- 10.7 keV), and because of its spectral resolution, 
which will exceed that of previous experiments at energies above 3 keV.  

Over the last several years, our group has worked on providing accurate K-shell 
atomic data for all ions of astrophysical interest.
We have calculated energy levels, wavelengths, Einstein $A$-coefficients,
Auger rates, and photoabsorption/ionization cross sections.
Complete data sets have been computed for all ion stages of Fe 
\citep{bautetal03,palmetal03,palmetal03b,mendetal04,bautetal04,kalletal04}, 
O \citep{garcetal05}, N \citep{garcetal09}, as well as Ne, Mg, Si, S, Ar,
and Ca \citep{palmetal08,wittetal09,wittetal11}.
There has also been recent work on the C sequence by \citet{hasoetal10} using 
similar techniques.

The data we present in this work are important both for their direct effect on 
observed astrophysical X-ray spectra via absorption features, and also through 
their effect on ionization balance calculations, which in turn affects many of 
the observables from these elements.
In particular, it is the detailed resonance and edge structure, missing in 
previous calculations, which is most important.

There have been a large number of calculations for these systems going back to
the 1920's, but they are confined to the non-resonant background and few span
the K edge region.
For the ions considered in this work, there is good agreement between our 
results and the background cross section both above and below the K edge 
characterized by the fits of \citet{vernyako95}.
Laboratory measurements have been mainly confined to neutral systems for
photon energies beyond the K edge.
A comprehensive assessment and fitting of measurements has been performed by
\citet{veig73}.

There has been recent progress measuring the resonance structure near the 
K edge of Li-like systems: C$^{3+}$ \citep{mulletal09} and 
B$^{2+}$~\citep{mulletal10}, as well as Be-like C$^{2+}$ 
\citep{sculletal05}.
$R$-matrix calculations performed as part of those works are in good agreement
with the measurements apart from small discrepancies in some resonance
positions and heights.
With the advances of free electron lasers to produce X-rays
\citep{kantetal06}, we expect it will be possible over the next
several years to perform similar experiments for the K-shell absorption by 
heavier species.
However, such measurements for Ni are not yet available.


\section{Numerical methods}

The $R$-matrix method employed here is the same as previous K-shell 
calculations in this series, c.f. \citet{bautetal03}.
The $R$-matrix method is based on the close-coupling approximation of
\citet{burkseat71}, which is solved numerically following 
\citet{burkhibbrobb71,berretal74,berretal78,berretal87}.

As in our previous calculations, the effect of radiative and spectator Auger
decay on resonances is included using the optical potential of 
\citet{gorcbadn96,gorcbadn00}. 
Here, the resonance energy with respect to the threshold acquires an imaginary
component which is the sum of the radiative and Auger widths of the core. 
The radiative widths are computed within the $R$-matrix calculations following
the prescription given by \citet{robietal95}. 
The Auger widths are taken from the relativistic Hartree-Fock calculations of
\citet{palmetal08b}.
These effects are referred to as radiative and Auger damping since they cause
resonances in the photoionization cross section to become smaller in area.
The damping also makes it necessary to calculate both the photoabsorption and 
photoionization cross sections since not every absorbed photon will result in a
photoelectron.

We use the $R$-matrix computer package of \cite{berreissnorr95} for the
inner region calculation and the asymptotic region code 
{\sc stgbf0damp}\footnote{http://amdpp.phys.strath.ac.uk/tamoc/code.html}
\citep{gorcbadn96} is used to compute the photoabsorption and photoionization 
cross sections including the effects of damping. 

For the Li-like to Ne-like ions, our target expansion includes all states in
the configurations: 1s$^2$ 2$l^w$ and 1s 2$l^{w+1}$ where $w=0$ for Li-like
and $w=7$ for Ne-like.  Note that the target refers to the final state which
has one fewer electron than the initial state; thus the Ne-like case having
a 9-electron target.
For the remaining ions, we must include $n=3$ orbitals.
The included configurations are 1s$^2$ 2$l^8$ 3$l^w$, 
1s$^2$ 2$l^8$ 3$l^{w-1}$ 3d, 1s$^2$ 2$l^7$ 3$l^{w+1}$, and 1s 2$l^8$ 3$l^{w+1}$
where $l$ is restricted to s and p electrons.
We only include 3d excited states from the valence configuration to improve the
atomic structure.  
We have tested the addition of 3d configurations in the L- and K-hole states
for Mg-like and Ar-like Ni and have found the following differences.
First, the $K\beta$ resonances are shifted by a couple eV which is within the
uncertainty of the calculation.
This difference in the resonance positions decreases approaching the K edge.
Secondly, the larger calculation has a few additional resonances, but these are
small and disappear when the cross section is convolved with a resolution
reflective of current X-ray detectors.
Therefore, we are confident in omitting the additional 3d configurations for
all of the present calculations.

We include enough continuum basis orbitals to span from threshold to at least
four times the energy of the K edge.
This choice gives reliable cross sections up to at least twice the energy
of the K edge.
Partial photoionization and total photoionization/absorption cross sections are
calculated from all states in the ground configuration of the parent ion.

The close-coupling expansion is small enough for the Li-like to Na-like ions
to allow for fully level-resolved, Breit-Pauli ({\sc bprm}) calculations to be 
carried out.
The number of fine-structure levels for the other ions quickly becomes too
large for the computational resources available, so we revert to $LS$-coupling 
for the remaining ions.
The mass-velocity and Darwin terms are still included in these calculations 
which greatly improve the target energies.
Wave functions are obtained using {\sc autostructure} \citep{badn86,badn97} 
where the level/term energies have been adjusted to match those in the NIST 
database \citep{ralckramread08} when possible, otherwise we match our energies
to those calculations by \citet{palmetal08b}.
Some levels in our target expansion are not listed by either of the above
sources.
For such valence and L-vacancy states, we use our {\sc autostructure} energies 
unshifted.
For the unlisted K-hole states, we shift our energies to match the average
difference between the K-hole states appearing in both our target and 
\citet{palmetal08b}.
This last shift is approximately 25 eV (0.3\%) for the Mg-like to Ca-like ions.
The resulting cross sections are shifted in energy so that the ionization
threshold matches NIST.  
This final shift brings the K edge thresholds within 0.5\% of those given by 
\citet{palmetal08b}.

Cross sections for Sc-like to Fe-like Ni are calculated using the distorted
wave method as implemented by {\sc autostructure}.
These calculations use the same configuration expansion as the $R$-matrix 
calculations with the exception of Fe-like Ni which includes the
3d$^6$ 4$\ell$ ($\ell = 0,1,2$) configurations.
Resonances are included using the isolated resonance approximation up to $n=20$.
These calculations are performed in $LS$ coupling and the energies are shifted
so that the location of the K edge matches \citet{palmetal08b}.
The background cross section is calculated up to 13.6 keV while resonances are
calculated only near the K edge.
{\sc autostructure} can construct wave functions from either the $N$-electron 
target orbitals or the $(N+1)$-electron parent orbitals
The difference in the cross sections using both sets of wave functions gives
an idea of the uncertainty of this method.
The wave functions are composed of relaxed (non-orthogonal) orbitals.
Since we do not attempt to take radiation or Auger dampings into account, we
only report total photoabsorption cross sections which are convolved with a
Gaussian profile.
The width of the Gaussian is $\Delta E/E = 10^{-3}$ which is represetative of
currently deployed X-ray detectors.
Calculations for Ni I and Ni II are more involved, so we leave these for 
future work.

\section{Results}

\begin{figure}
\includegraphics[width=10cm,angle=270]{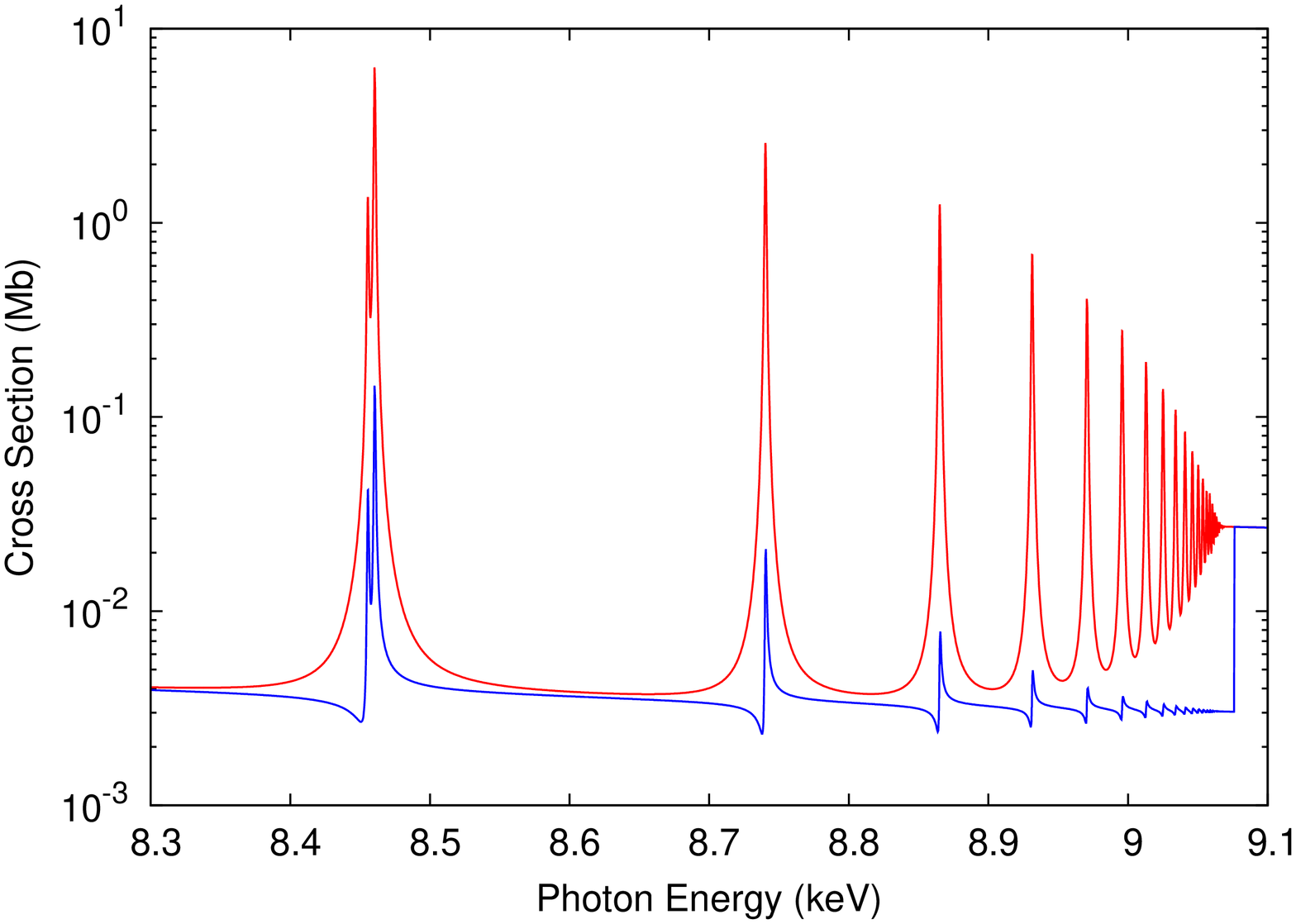}
\caption{
Photoabsorption (upper, red) and photoionization (lower, blue) cross sections 
of Ne-like Ni near the K edge.  Color available in the on-line version of the 
text.
}
\label{fig:ni.nelike}
\end{figure}

The cross sections reported here contain many of the same features shown in
the results from our previous calculations.  
Inclusion of radiation and Auger damping broadens resonances near the K edge
and makes them more symmetric.
The damping is also important in distinguishing the absorption and ionization 
cross sections.
In Figure~\ref{fig:ni.nelike}, we show these cross sections for the $^1$S$_0$ 
ground state of Ne-like Ni.
As there are no K$\alpha$ resonances for Ne-like ions, the first resonance is
K$\beta$. 
The effect of damping is observed in both the photoionization and 
photoabsorption resonances.
In the former, the heights of the resonances are greatly diminished,
especially as the photon energy approaches the K edge due to the dominance
of the spectator Auger process.
For photoabsorption, both radiation and Auger damping broaden the resonances
and make them more symmetric.

\begin{figure}
\includegraphics[width=10cm,angle=270]{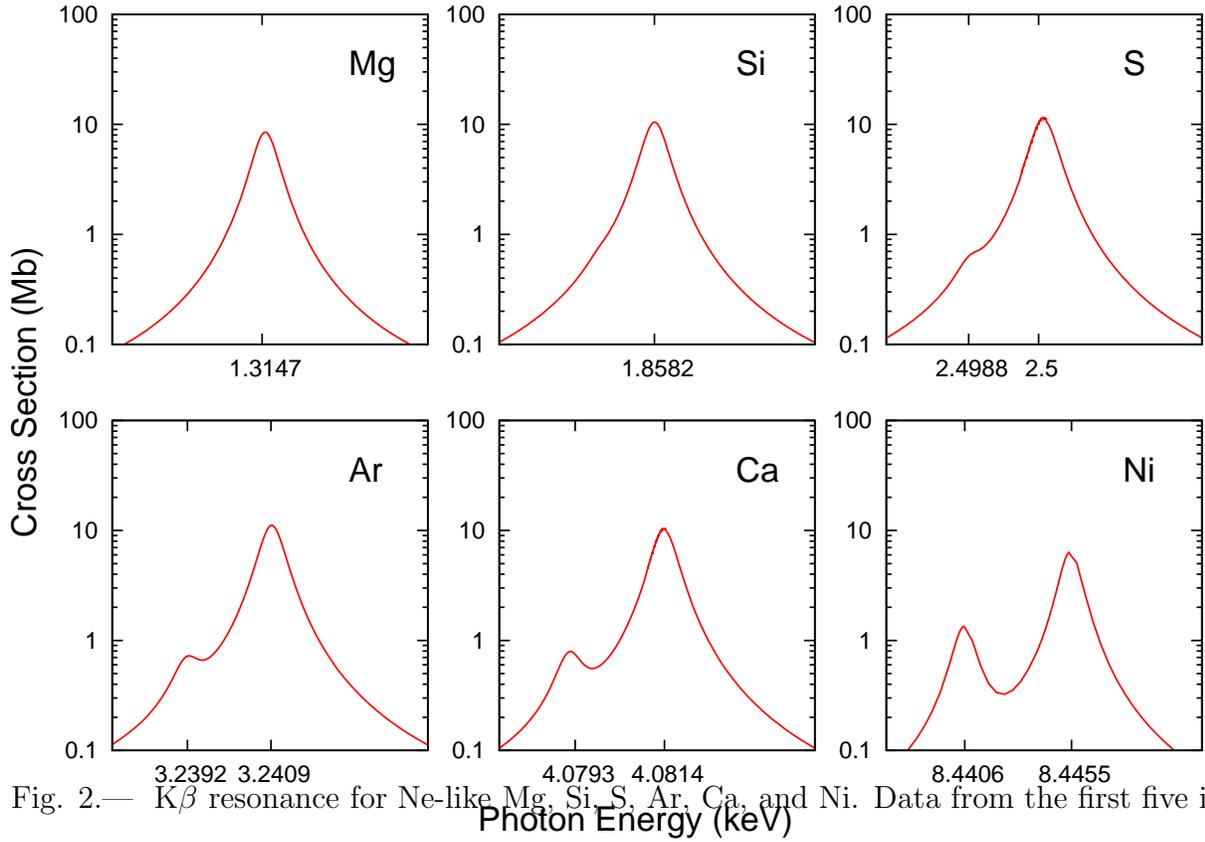}
\caption{
K$\beta$ resonance for Ne-like Mg, Si, S, Ar, Ca, and Ni.  
Data from the first five ions comes from \citet{wittetal09}.
The second resonance appearing from Ne-like S to Ni is due to photoexcitation 
to the 1s 2s$^2$ 2p$^6$ 3p $^3$P$_1$ level.
}
\label{fig:ni.nelike.kbeta}
\end{figure}

For this simple system, one would expect a single resonance for each Rydberg
$n$-shell, however, the K$\beta$ resonance has two peaks in our results.
This represents the breakdown of $LS$ coupling at large $Z$.
For low-$Z$ ions, the 2p$^6$ $^1$S$_0$ ground state goes to the 
1s 2s$^2$ 2p$^6$ 3p $^1$P$_1$ state.
At high $Z$, the total spin is no longer a conserved quantity, making the
$^3$P$_1$ state accessible leading to the second K$\beta$ peak.
For the $n>3$ resonances, $LS$ coupling remains a good description for a Rydberg
electron attached to a core, so only a single resonance is found.
The appearance of the second K$\beta$ resonance as $Z$ increases is 
demonstrated in Fig~\ref{fig:ni.nelike.kbeta} where we show the K$\beta$ 
resonance for six Ne-like ions from $Z=12$ to 28 using data from 
\citet{wittetal09}.
The separation of the K$\beta$ resonances for Ne-like Ni is approximately 5 eV 
which should be detectable in the next generation of X-ray observatories, such 
as {\it Astro-H}.

\begin{figure}
\includegraphics[width=10cm,angle=270]{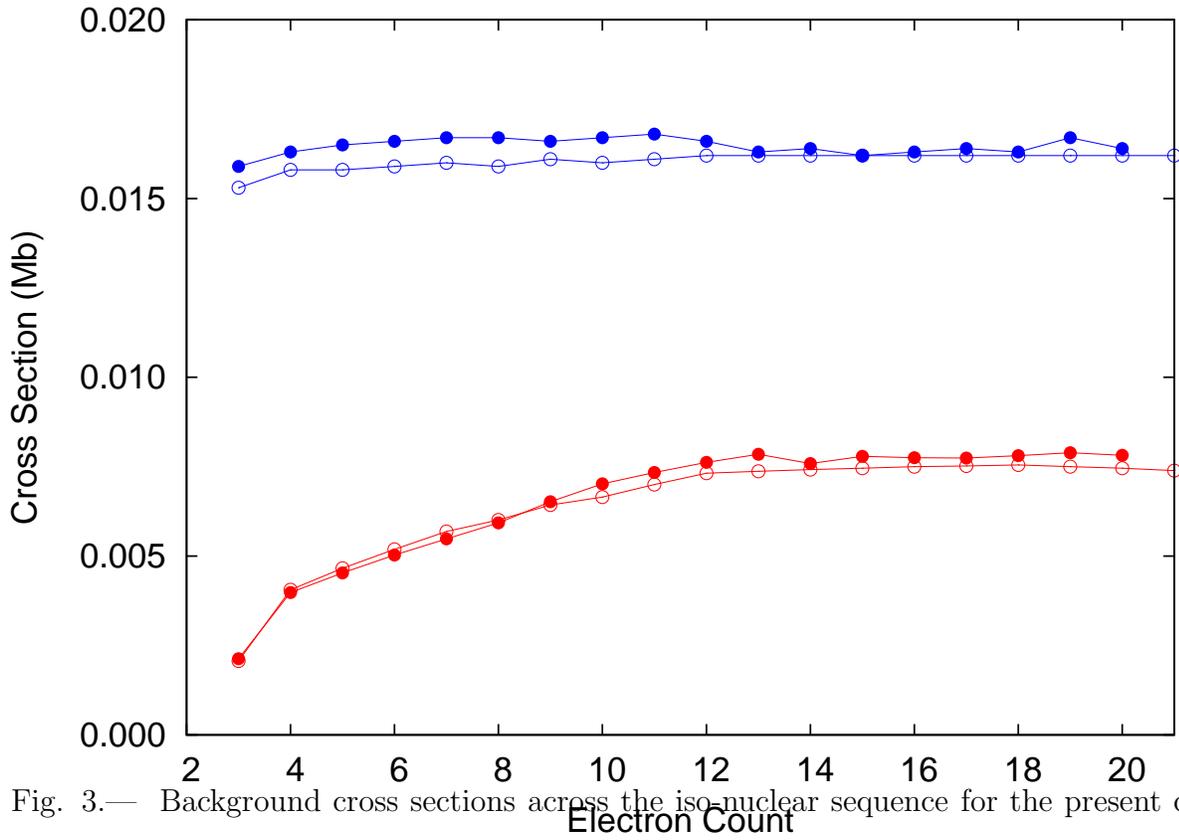}
\caption{
Background cross sections across the iso-nuclear sequence for the present 
calculations (filled circles) and \citet{vernyako95} (open circles).
The upper two curves (blue) show the cross section above the K edge at
10.9 keV; lower two curves (red) give the cross section below the K
edge at 6.8 keV.  Color available in the on-line version of the 
text.
}
\label{fig:trend}
\end{figure}

In Figure~\ref{fig:trend}, we show the background cross sections at fixed 
photon energies above and below the K edge (at 10.9 keV and 6.8 keV, 
respectively) for all Ni ions included in this work.
For comparison, we also include the results from \citet{vernyako95}.  
Both results agree within a few percent along the whole iso-nuclear 
sequence.
The background below the K edge steadily increases until the $n=2$ shell is
filled (Ne-like), where the background becomes constant.
This is because the background is dominated by the L-shell cross section 
and the 2s and 2p electron wave functions are not affected much by the addition 
of $n=3$ electrons.
For the same reason, but for 1s electrons, the background cross section above 
the K edge is relatively constant beyond He-like Ni.

\begin{figure}
\includegraphics[width=10cm,angle=270]{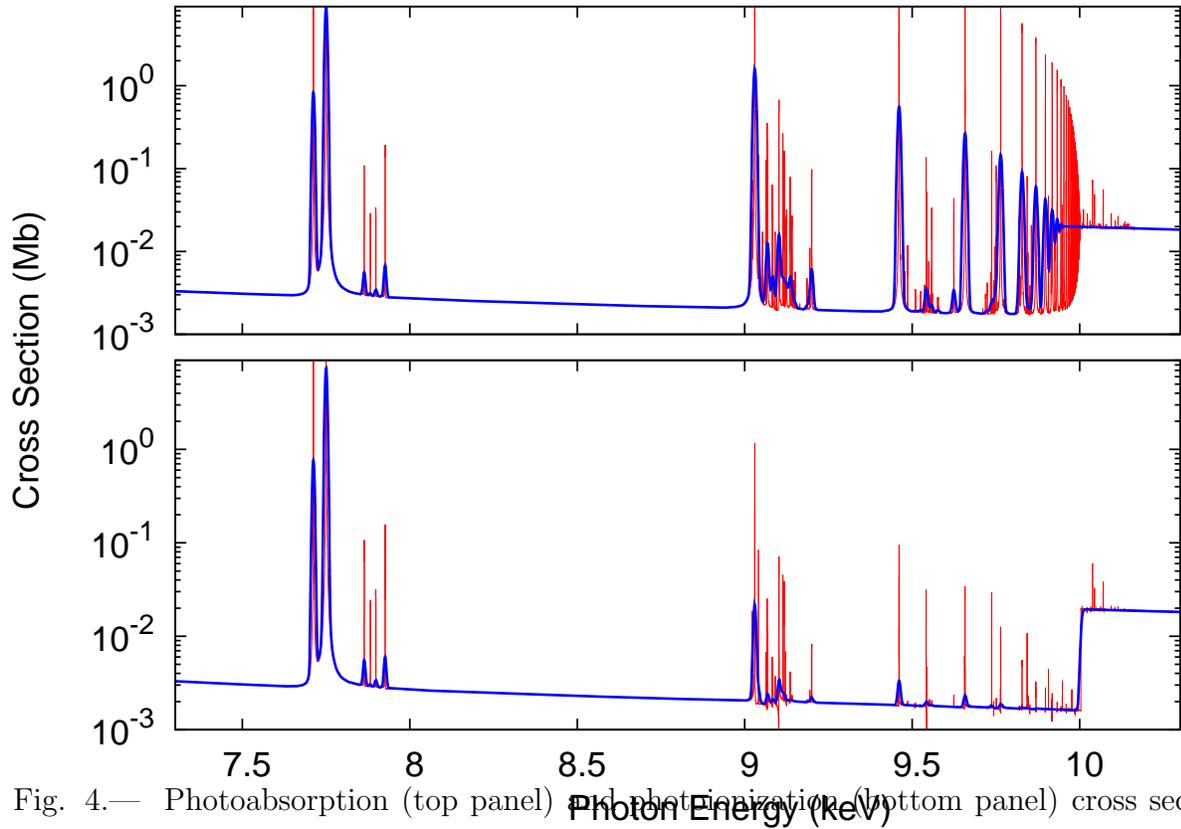}
\caption{
Photoabsorption (top panel) and photoionization (bottom panel) cross sections
for Be-like Ni.  Red, thin curves show the raw cross section and the blue, 
thick curves give the convolution with a Gaussian having a width of 
$\Delta E/E$ = 10$^{-4}$.  Color available in the on-line version of the 
text.
}
\label{fig:convolve}
\end{figure}

Resonances in the cross section are often quite narrow, even when Auger damping
is included, requiring tens of thousands of energy points to resolve all the 
features.
In modeling codes, this amount of data is cumbersome, so we convolve our cross
sections with a Gaussian profile having a width of $\Delta E/E = 10^{-4}$.
An example of the convolution is shown in Figure~\ref{fig:convolve} for Be-like
Ni for both photoabsorption (top panel) and photoionization (bottom panel).
The convolved results demonstrate the strength of damping as the resonant
contribution to the cross section for photoionization is nearly gone at the K
edge.
For photoabsorption, on the other hand, the resonances converge onto the K edge,
causing the perceived edge in the convolved results to occur at a smaller 
energy; this is known as the smearing of the K edge in absorption spectra.

\begin{figure}
\includegraphics[width=10cm,angle=270]{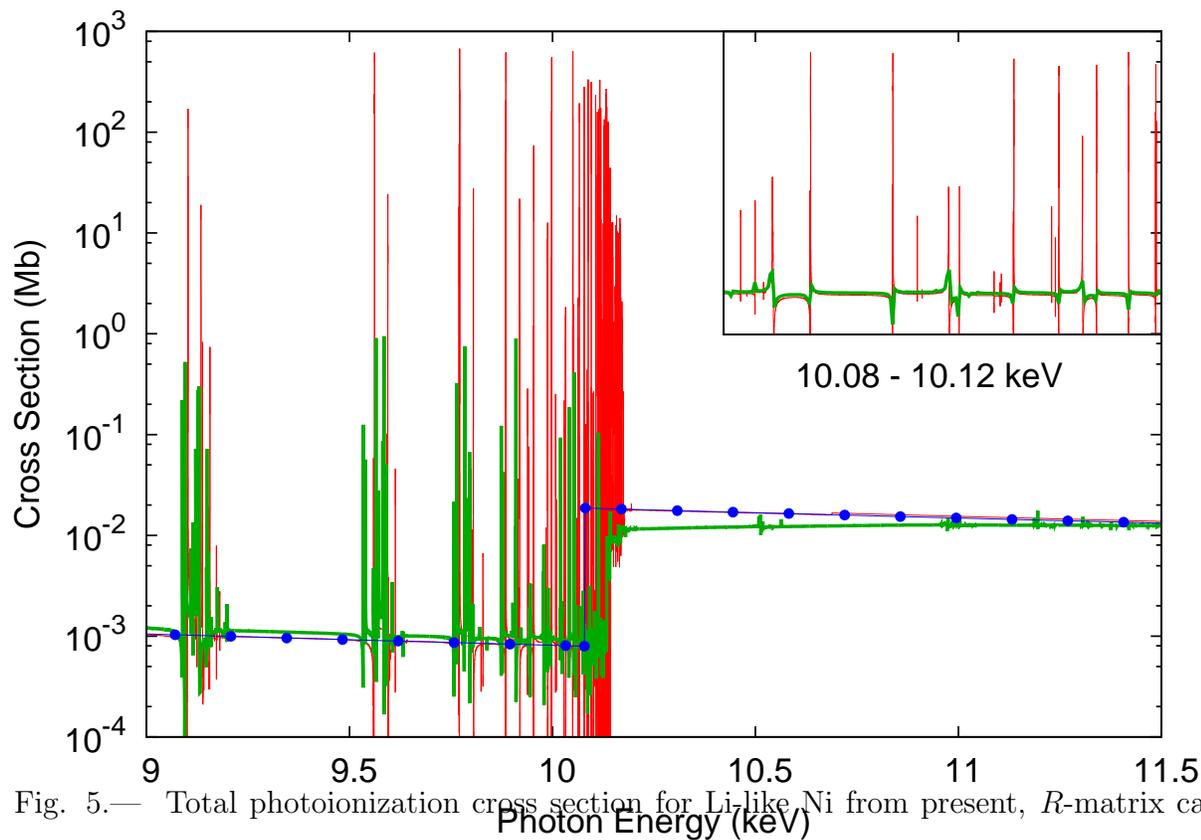}
\caption{
Total photoionization cross section for Li-like Ni from present, $R$-matrix
calculation (thin, red curve), the $R$-matrix calculation of \citet{naha05} 
(thick, green curve), and \citet{vernyako95} (thin, blue curve with circles).
The K$\alpha$ resonances are not visible in this energy range.  The background
cross sections of \citet{vernyako95} are in excellent agreement with the 
present work.  Inset: close-up of cross sections between 10.08 and 10.12 keV; 
range of cross sections is from 10$^{-4}$ to 10$^3$.  In the inset only, the 
\citet{naha05} cross sections have been shifted by 8.8 eV to align the resonance
positions.  Color available in the on-line version of the text.
}
\label{fig:nahar}
\end{figure}

As part of an effort to calculate recombination rate coefficients, 
\citet{naha05} performed a Breit-Pauli $R$-matrix calculation for Li-like
Ni with energies extending beyond the K edge.
We are in good agreement with those results for the ground state with respect
to the resonance positions and the location of the K edge.
However, there is some disagreement with the background cross section in the 
vicinity of the edge.
In Figure~\ref{fig:nahar} the present results are compared with the results of 
\citet{naha05} and \citet{vernyako95}.
Below the edge, we find fair agreement, the background of Nahar being about 
10\% higher than the present results, while above the edge, the Nahar 
background is nearly 30\% lower.
The disagreement above the edge disappears entirely above a photon energy of
11.5 keV.
The background from the present results are in excellent agreement with 
\citet{vernyako95} at all energies, although their K edge position is about 
50 eV lower than that calculated by ourselves and \citet{naha05}.

We find larger differences with \citet{naha05} in the amount of resonant 
enhancement.
This appears to be due to the higher energy resolution used in our calculation.
Unlike the rest of the ions covered in this work, the Li-like system does not
have a spectator Auger process which would significantly damp the narrow 
photoionization resonances near the K edge, as observed in 
Figure~\ref{fig:convolve}.
Instead, there is a large number of extremely narrow, yet strong, resonances 
which need to be fully resolved to account for their area.
\citet{naha05} uses about 20000 energy points to map out the cross section.
For the present calculation, we use the technique described in
\citet{wittetal09} which automatically identifies the location of each resonance
and adds enough energy points to fully resolve it.
Even with the effeciency of this approach, over a half million points were 
required to properly resolve all resonance features.
To illustrate the strength of the narrow resonances near the K edge, we show
in the inset of Figure~\ref{fig:nahar} the resonance structure in a narrow 
energy range.  
The difference between the calculations becomes particularly apparent when
comparing convolved cross sections.
Using a Gaussian width of $\Delta E/E = 10^{-3}$, we find that the 
\citet{naha05} K$\alpha$ resonances are about 5 times weaker than in the present
results; near the K edge, the difference is nearly an order of magnitude. 
These differences are expected to be important for spectral modeling of X-ray
plasmas.

\begin{figure}
\includegraphics[width=10cm,angle=270]{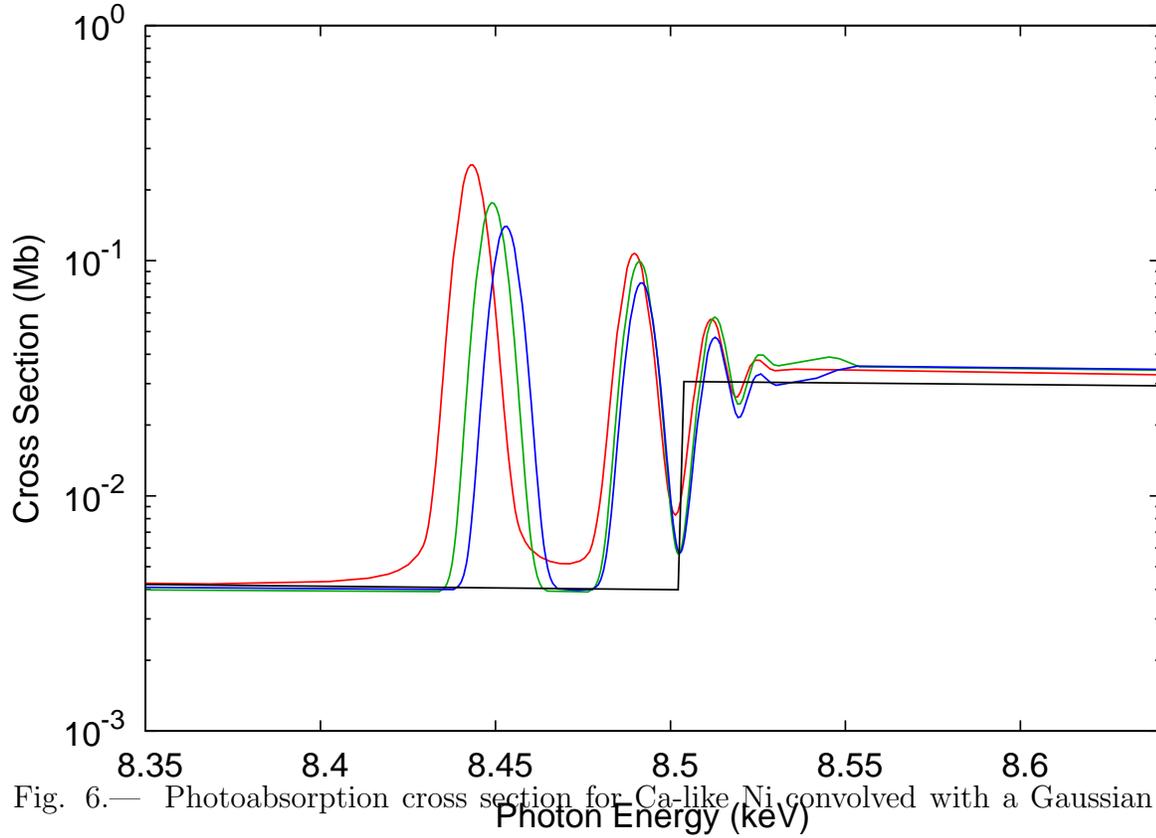}
\caption{
Photoabsorption cross section for Ca-like Ni convolved with a Gaussian having
a width of $\Delta E/E = 10^{-3}$.  Based on the height of the resonance
feature near 8.45 keV, in order from strongest to weakest, the curves are:
$R$-matrix results (red), {\sc autostructure} using target wave functions
(green), {\sc autostructure} using parent wave functions (blue), and the 
background cross section of \citet{vernyako95} (black).  Color available in 
the on-line version of the text.
}
\label{fig:calike_autos}
\end{figure}

Total photoionization cross sections have been calculated using 
{\sc autostructure} for Ar-like to Fe-like Ni.  
The first three ions of this sequence have also been calculated using 
$R$-matrix which allows for a comparison of the methods.
In Figure~\ref{fig:calike_autos}, we compare the present $R$-matrix results 
with two sets of {\sc autostructure} results using wave functions constructed
from target or parent orbitals.
All data have been convolved with a Gaussian using a width of
$\Delta E/E = 10^{-3}$.
Also shown are the direct photoionization cross sections of \citet{vernyako95}.
The {\sc autostructure} results have been shifted in energy ($\sim25$ eV) so 
that the K edge is in agreement with \citet{palmetal08b}.
The \citet{vernyako95} results are unshifted and a disagreement of $\sim50$ eV
in the edge position is observed.
The background cross sections above and below the K edge from the $R$-matrix
and both {\sc autostructure} calculations are in good agreement.
The background of \citet{vernyako95}, on the other hand, is about 10\% lower
than the $R$-matrix results just above the K edge, but comes into good
agreement by 10 keV.
There is some disagreement in the position and height of the first resonance 
(K$\gamma$) which is due to the relaxation effects on the orbitals in the 
{\sc autostructure} calculations which are not included in $R$-matrix.
This disagreement becomes smaller for the rest of the resonance series.
If we were to use orthogonal orbitals in the {\sc autostructure} calculations,
the agreement of the resonance positions with $R$-matrix would improve, but the
discrepancy of the resonance heights remains.
Since the {\sc autostructure} calculation using the target orbitals gives
better agreement with $R$-matrix than that using parent orbitals (which is 
also true for Ar-like and K-like Ni), we continue to use target orbitals for
the Sc-like to Fe-like {\sc autostructure} calculations.
Clearly these systems would benefit from future $R$-matrix calculations, but 
in the meantime, the {\sc autostructure} data provide an improvement over
previous calculations where resonances are neglected entirely.

By default, {\sc autostructure} calculates cross sections using the length 
gauge.  
This begins to fail numerically at high energies above the K edge and the 
velocity gauge is more accurate.
We take cross sections calculated using the length-gauge at energies below the
K edge and the velocity gauge results for energies above the edge.
This method yields a background cross section in good agreement with the
$R$-matrix results for Ar-like to Ca-like Ni, and is applied to the 
other systems.

Data from this work are to be used in the {\sc xstar} code
\citep{kallbaut01,bautkall01} for modeling photoionized plasmas.
To prepare for this application, we need to reduce the number of data points
from thousands to hundreds.
In addition to convolution (as described earlier), we remove unneccessary data 
points from each data set in the following manner.
Points which lie on a straight line between the two adjacent points within 1\%
are removed.
This test is repeated over the entire cross section until no more points are
removed.
Using this process, we can decrease the number of points in the cross section
data from several thousand to a few hundred, yet still maintain good 
accuracy with linear interpolation.
The convolved cross sections are not included with this paper, but are
available by a request to MCW, 
{\sc xstar}\footnote{http://heasarc.gsfc.nasa.gov/xstar/xstar.html}, or the 
Universal Atomic Database\footnote{http://heasarc.gsfc.nasa.gov/uadb}.
The raw cross sections are available as electronic tables attached to this work.

The energy range of all datasets in the present work begin at the valence-shell
threshold.
However, as we are only focused on the resonance features near the K edge, we
have not attempted to fully resolve the resonances in the region of the 
L edge.


\section{Summary and Conclusions}

Total photoabsorption and photoionization cross sections have been computed
for the Li-like to Fe-like ions of Ni.
Level-resolved, Breit-Pauli $R$-matrix calculations were carried out for the
Li-like to Na-like ions.
Term-resolved $R$-matrix calculations were performed for Mg-like to Ca-like
where the mass-velocity and Darwin relativistic corrections are included.
Distorted wave calculations for Sc-like to Fe-like Ni were performed using
{\sc autostructure} where resonances are included using the isolated resonance
approximation.

For all of the $R$-matrix calculations, total photoabsorption and total/partial
photoionization data are available.
Radiation and Auger dampings are included to account for radiation losses and 
the spectator Auger process.
No damping is included for the {\sc autostructure} calculations so we present
only total photoabsorption cross sections which are convolved with a 
Gaussian profile using a width representative of currently deployed X-ray
detectors.

We find overall good agreement with the background cross sections given by
\citet{vernyako95}, but our positions of the K edge are typically 40-80 eV
higher.
For Li-like Ni, we are in good agreement with \citet{naha05} as to resonance
positions. 
However, we find much stronger resonant enhancement due to the finer energy
mesh used in the present work.

All data are provided as on-line tables accompanying this article and can also
be obtained through the {\sc xstar} atomic database \citep{bautkall01} 
and the Universal Atomic Database.
The data sets provided here together with the energy levels and radiative and 
Auger rates reported in \citet{palmetal08b} will help modelers to carry
out detailed studies of K spectra of Ni ions.

\begin{acknowledgments}
Support for this research was provided in part by a grant from the NASA 
Astronomy and Physics Research (APRA) program.
PP and PQ are respectively Research Associate and Senior Research Associate
of the Belgian F.R.S.-FNRS.  
Financial support from this organization is acknowledged.
\end{acknowledgments}


\bibliography{mybib}


\end{document}